\hfuzz 2pt
\vbadness 10000
\font\titlefont=cmbx10 scaled\magstep1

\magnification=\magstep1

\def\asymptotic#1\over#2{\mathrel{\mathop{\kern0pt #1}\limits_{#2}}}

\null
\vskip 1cm
\centerline{\titlefont COMPLETE POSITIVITY}
\medskip
\centerline{\titlefont AND ENTANGLED DEGREES OF FREEDOM}
\vskip 2cm

\centerline{\bf F. Benatti}
\smallskip
\centerline{Dipartimento di Fisica Teorica, Universit\`a di Trieste}
\centerline{Strada Costiera 11, 34014 Trieste, Italy}
\centerline{and}
\centerline{Istituto Nazionale di Fisica Nucleare, Sezione di
Trieste}
\vskip .5cm

\centerline{\bf R. Floreanini}
\smallskip
\centerline{Istituto Nazionale di Fisica Nucleare, Sezione di
Trieste}
\centerline{Dipartimento di Fisica Teorica, Universit\`a di Trieste}
\centerline{Strada Costiera 11, 34014 Trieste, Italy}
\vskip .5cm

\centerline{\bf R. Romano}
\smallskip
\centerline{Dipartimento di Fisica Teorica, Universit\`a di Trieste}
\centerline{Strada Costiera 11, 34014 Trieste, Italy}
\centerline{and}
\centerline{Istituto Nazionale di Fisica Nucleare, Sezione di
Trieste}
\vskip 2cm

\centerline{\bf Abstract} 
\smallskip
\midinsert
\narrower\narrower\noindent
We study how some recently proposed noncontextuality tests based on 
quantum interferometry are affected if the test particles
propagate as open systems in presence of a gaussian stochastic
background.
We show that physical consistency requires the resulting markovian dissipative 
time-evolution to be completely positive.
\endinsert
\bigskip
\vfill\eject

\noindent
{\bf 1. INTRODUCTION}
\medskip

Recently, experiments based on neutron [1] and photon [2] interferometry
have been proposed to test the hypothesis of noncontextuality in 
quantum mechanics; the idea is to check whether
a Bell-like inequality of the Clauser-Horne-Shimony-Holt form [3] 
is violated or not.
Such an inequality is derived from the assumption that the measured values of
physical observables are completely specified by the state of the system 
prior to measurement and that the actual measurement outcomes  
do not depend on the context, namely on whether other commuting observables 
are simultaneously measured.

Differently from Bell-locality tests based on entangled 
physical systems, the above experiments involve two degrees of 
freedom of a same physical system;
one degree of freedom is translational, related to
the two possible paths followed by
neutrons or photons inside the 
interferometer, the other 
is the spin (helicity) of neutrons (photons).

Standard Bell-locality tests are not concerned with the
time-evolution of the particles involved; only when there is lack of 
unitarity and loss of probability as in experiments based
on decaying neutral K-mesons [4-6], the time-evolution becomes
important.

More in general, the dynamics is to be taken into account when
the test particles behave as open systems $S$ propagating through an 
environment $\cal E$ to which they are coupled;
in such cases, one usually traces away the degrees of freedom of
$\cal E$ ending up, under certain assumptions, with a one-parameter
semigroup of linear maps $\Gamma_t$ 
on the states of $S$ represented by density matrices $\rho$.

The maps $\Gamma_t$ constitute a so-called {\it reduced dynamics} for the 
open quantum system $S$ and embody the dissipative and mixing effects due 
to the environment $\cal E$; they are not unitary, satisfy the forward in time 
composition law
$\Gamma_{t+s}=\Gamma_{t}\circ\Gamma_{s}$ for 
$s,t\geq0$, and  transform 
pure states into statistical
mixtures.
When they enjoy the property known as
complete positivity they form a so-called  
``quantum dynamical semigroup'' [7-10].

Physical consistency requires that the positivity of states of $S$,
that is the positivity of the
eigenvalues of the corresponding density matrices, be preserved for all times
by any meaningful reduced dynamics $\Gamma_t$; 
indeed, the statistical interpretation of
quantum mechanics identifies these eigenvalues with probabilities.

The property of complete positivity guarantees not only that
the maps $\Gamma_t$ preserve the positivity of the states of $S$, but also
that the maps
${\bf I}_N\otimes\Gamma_{t}$
preserve the positivity 
of all states of
the composite system $S_{N}+S$, for any $N$-level system, with
${\bf I}_{N}$ the 
identity operation on $S_N$ [11,12].
Complete positivity of $\Gamma_{t}$ is stronger than positivity
and is intimately connected with quantum entanglement;
indeed, positivity alone  is not sufficient to ensure that 
${\bf I}_{N}\otimes\Gamma_{t}$ preserve the positivity of entangled 
states of $S_{N}+S$.

Noticeably, the standard quantum mechanical time-evolution generated by
Hamiltonian operators is unitary, reversible and
completely positive.
On the contrary, the physical
literature abounds with dissipative, irreversible, reduced dynamics of quantum
open systems that are 
neither positive, nor completely positive, see {\it e.g.} [13-15].

In particular, in view of the abstract, experimentally uncontrollable
coupling of the system of interest $S$ with {\it any} $N$-level system $S_N$,
the argument that $\Gamma_t$ should necessarily be 
completely positive may look as a mathematical convenience and 
a technical artifact, rather than a physical necessity [16].
In fact, the elimination of the environment degrees of freedom yields 
an equation of motion with memory terms that have to be
eliminated via suitable Markov approximations.
It depends on how they are performed whether the resulting 
semigroups are physically consistent or not [17].

Recently, the issue of complete positivity has been reconsidered in
the context of neutral meson dynamics, where a typical experimental
situation is that of an
entangled, singlet-like, state $\rho$ of two K or B neutral mesons 
propagating according to the
factorized time-evolution $\Gamma_t\otimes\Gamma_t$ [18-21].
If the dynamical maps $\Gamma_t$ are assumed to be not of the standard 
Weisskopf-Wigner form, but modified by a noisy 
background of gravitational origin, it is showed that $\Gamma_t$ has to be
completely positive.
Were it not so, physical inconsistencies as the production of negative
probabilities, would affect $\Gamma_t\otimes\Gamma_t[\rho]$; moreover,
these inconsistencies cannot be dismissed as experimentally invisible 
because they might give rise to 
detectable effects [20].

In this paper we consider physical cases where the
environment $\cal E$ is given by a classical, fluctuating external field 
[13,22] and $S$ is a single open quantum system with two degrees of freedom.
In such a context, the coupling is not between $S$ and an abstract
$2$-level system, but
between two degrees of freedom of $S$ itself.
Then, 
the physical meaning of complete positivity comes to the 
fore when we study a time-evolution of the form 
${\bf I}_N\otimes\Gamma_t$.

Namely, we study what happens if
the interferometric apparatuses proposed for noncontextuality tests 
are placed
in weak gaussian stochastic magnetic fields, or stochastic optical media, 
coupled to the spin, or helicity of
neutrons, respectively photons,
that will then propagate as open quantum systems.

In particular, in the case of neutrons,
we will consider in detail three possible choices of 
fluctuating magnetic backgrounds; these will give rise to reduced dynamics
${\bf I}_2\otimes\Gamma_t$ with $\Gamma_t$ affecting the neutron
spin degree of freedom covering all possible cases, namely to
$\Gamma_t$ completely positive, positivity preserving, but not completely
positive and, finally, to $\Gamma_t$ not even positivity preserving.
The three possibilities depend on the properties of the stochastic magnetic
field; it thus appears that by reproducing stochastic magnetic
backgrounds with the qualities of the three cases we are referring to,
one would be able to experimentally study the characteristics of the various
reduced dynamics.

Further, we study how the Clauser-Horne-Shimony-Holt inequality 
is modified by the stochastic magnetic field
and show that   
complete positivity of the time-evolution ${\bf I}_{N}\otimes\Gamma_{t}$
is necessary for a consistent physical 
description; otherwise
unacceptable negative eigenvalues appear in the time-evolving
physical states describing entangled degrees of freedom.

\vskip 1cm

\noindent
{\bf 2. ENTANGLEMENT AND NON-CONTEXTUALITY TESTS}
\medskip

In the following we shall refer to noncontextuality tests using
neutron interferometry [1],
photons involving similar arguments [2].
In the experimental setup proposed in [1], 
an incoming beam of neutrons with spin along the positive $z$-direction
passes through a beam splitter with 
transmission and reflection coefficients $p$ and $q$ with
$|p|^2+|q|^2=1$.
The beam gets divided into two components that 
follow two spatially separated paths $u$ and $d$.

Both the spin and the translational degree of freedom are described
by $2$-dimensional Hilbert spaces, the former
with basis vectors $|\uparrow_z\rangle$ and
$|\downarrow_z\rangle$, 
the latter with basis vectors $|\psi_u\rangle$ and $|\psi_d\rangle$ 
corresponding to
the two possible macroscopic paths.
Making the beam $u$-component undergo a
spin-flip $|\uparrow_z\rangle\mapsto|\downarrow_z\rangle$, 
an initial beam state is prepared,
$$
|\Psi\rangle=p\,|\psi_u\rangle\otimes|\downarrow_z\rangle\ +\
q\,|\psi_d\rangle\otimes|\uparrow_z\rangle
\ ,\eqno(2.1)
$$ 
which then propagates inside the interferometer.
A vector state as above corresponds to the one-dimensional projector 
$\rho_{\Psi}:=|\Psi\rangle\langle\Psi|$, 
$$
\rho_\Psi=
|p|^2P_1\otimes Q_2\,+\,|q|^2P_2\otimes Q_1
\,+\,pq^*\,P_3\otimes Q_4\,+\,p^*q\,P_4\otimes Q_3\ ,
\eqno(2.2)
$$
with
$$
\eqalignno{P_1&:=|\psi_u\rangle\langle\psi_u|\ ,\
P_2:=|\psi_d\rangle\langle\psi_d|\ ,\
P_3:=|\psi_u\rangle\langle\psi_d|\ ,\
P_4:=|\psi_d\rangle\langle\psi_u|\ ,&(2.3{\rm a})\cr
\cr
Q_1&:=|\uparrow_z\rangle\langle\uparrow_z|\ ,\
Q_2:=|\downarrow_z\rangle\langle\downarrow_z|\ ,\
Q_3:=|\uparrow_z\rangle\langle\downarrow_z|\ ,\
Q_4:=|\downarrow_z\rangle\langle\uparrow_z|\ .&(2.3{\rm b})}
$$
More in general, neutron beam states are not pure as $\rho_\Psi$, rather
statistical mixtures described by density matrices 
$$
\rho_{(2)}:=\sum_{i,j=1}^4\rho_{ij}\, P_i\otimes Q_j\ ,
\eqno(2.4)
$$
that is by $4\times 4$ hermitian, normalized, positive 
matrices, whose positive eigenvalues 
sum up to one (${\rm Tr}\rho_{(2)}=1$).
Positivity is crucial for the statistical interpretation of
quantum mechanics where the eigenvalues of density matrices
play the role of probabilities.

Neutrons spend a typical time $t$ within the interferometer
during which they may be subjected to external influences resulting
in a dynamical change $\rho_{(2)}\longmapsto\rho_{(2)}(t)$ of their state.
At the exit of the interferometer,
a second beam splitter recombines the translational components
and shifts the $u$ component by an angle $\varphi$,
$$
\pmatrix{
|\psi_u\rangle\cr
|\psi_d\rangle}\longmapsto\pmatrix{
|\psi_u(\vartheta,\varphi)\rangle\cr
|\psi_d(\vartheta,\varphi)\rangle }=
\pmatrix{
{\rm e}^{-i\varphi}\sin\vartheta&\cos\vartheta\cr
{\rm e}^{-i\varphi}\cos\vartheta&-\sin\vartheta}
\pmatrix{
|\psi_u\rangle\cr
|\psi_d\rangle} \ ,
\eqno(2.5)
$$
with reflection and transmission probabilities 
$\cos^2\vartheta$ and $\sin^2\vartheta$.
Consequently, the neutron beam state emerging from the interferometer
is
$$
\rho\longmapsto
(U(\vartheta,\varphi)\otimes{\bf 1}_2)\,\rho_{(2)}(t)\,
(U^*(\vartheta,\varphi)\otimes{\bf 1}_2)\ ,
\eqno(2.6)
$$
where $U(\vartheta,\varphi)$ is 
the unitary matrix in (2.5), 
$U^*(\vartheta,\varphi)$ its adjoint and ${\bf 1}_2$ is the $2\times 2$ 
identity matrix.
The two components of the exiting beam are then
intercepted 
by two counters $C_{u,d}$ plus
spin-analyzers $S_{{\bf n},-{\bf n}}$ that 
record how many neutrons reach them with
spins polarized along suitable directions $\pm{\bf n}=\pm(n_1,n_2,n_3)$
in space.
The frequencies of counts give the expectations
$$
{\cal O}^{j,{\bf n}}_t(\vartheta,\varphi)
:={\rm Tr}\Bigl(\rho_{(2)}(t)\, P_j(\vartheta,\varphi)
\otimes Q_{\bf n}\Bigr)\ ,
\eqno(2.7)
$$
where $P_j(\vartheta,\varphi):=U^*(\vartheta,\varphi)\,P_j\, 
U(\vartheta,\varphi)$,
$P_j$, $j=1,2$, are as in (2.3a) and represent observables that are chosen by
setting the angles $\vartheta$, $\varphi$ characteristic of the beam splitter.
Further, $Q_{{\bf n}}:=
|\uparrow_{{\bf n}}\rangle\langle\uparrow_{{\bf n}}|$ projects onto a 
state with
spin along the direction ${\bf n}$.

Since translational and spin observables commute, 
the observables 
$$
A(\vartheta,\varphi):=P_1(\vartheta,\varphi)\ -\ P_2(\vartheta,\varphi)\ ,
\quad
B({\bf n}):=Q_{{\bf n}}-Q_{-{\bf n}}\eqno(2.8)
$$
also commute and have eigenvalues $\pm1$.
Choosing angles $\vartheta_{1,2}$, $\varphi_{1,2}$ and polarization 
directions
${\bf n}_{1,2}$, one constructs commuting observables 
$A_i$, $B_j$, $i,j=1,2$, called dichotomic~[1].

In the hypothesis of noncontextuality, the possible outcomes $\pm1$
of a measurement of $A_i$, 
respectively $B_j$, are predetermined by the state $\rho_{(2)}(t)$
independently on whether $B_j$, respectively $A_i$, 
are simultaneously measured with $A_i$, respectively $B_j$.
From such assumptions a Clauser-Horne-Shimony-Holt inequality
can be derived for the mean values
$$
\eqalign{
C_t(\vartheta,\varphi;{\bf n})&:=
{\rm Tr}\Bigl(\rho_{(2)}(t)A(\vartheta,\varphi)\otimes B({\bf n})\Bigr)\cr
&={\cal O}^{1,{\bf n}}_t(\vartheta,\varphi)+
{\cal O}^{2,-{\bf n}}_t(\vartheta,\varphi)-
{\cal O}^{1,-{\bf n}}_t(\vartheta,\varphi)-
{\cal O}^{2,{\bf n}}_t(\vartheta,\varphi)\ ,}
\eqno(2.9)
$$
with four possible configurations of the control parameters 
$\vartheta$, $\varphi$ and ${\bf n}$,
$$
\Bigl|C_t(\vartheta_1,\varphi_1;{\bf n}_1)+
C_t(\vartheta_1,\varphi_1;{\bf n}_2) +
C_t(\vartheta_2,\varphi_2;{\bf n}_1)-
C_t(\vartheta_2,\varphi_2;{\bf n}_2)\Bigr|\leq 2\ .
\eqno(2.10)
$$ 
From (2.9), it turns out that the quantities $C_t(\vartheta,\varphi;{\bf n})$ 
can be measured by frequencies
at counters plus spin-analyzers $(C_u,S_{{\bf n}})$ and 
$(C_d,S_{-{\bf n}})$ 
with the beam splitter at the exit of the interferometer
set at angles $\vartheta$ and $\varphi$.

If inequality (2.10) is violated, the hypothesis of
noncontextuality upon which it was derived, cannot hold.
Whether it is so or not can be checked in highly
efficient experiments where the
entanglement between translational and magnetic degrees of freedom 
is exploited [1]
(for a similar argument involving photons see [2]).

Interestingly, by setting appropriately the angles $\vartheta$ and $\varphi$
of the beam-splitter and the polarization direction ${\bf n}$ of 
the spin-analyzers, also the
entries of the state 
$\rho_{(2)}(t)=\sum_{i,j}\rho_{ij}(t)P_i\otimes Q_j$, 
can be measured.
From the entries, one has access to the eigenvalues of the beam 
state after travelling through the interferometer and thus to
the effects of the time-evolution inside it.
Indeed, from (2.3, 2.4) and (2.6) it readily follows that  
$$
\rho_{11}(t)={\cal O}^{1,z}_t(0,0)\ ,\
\rho_{12}(t)={\cal O}^{1,-z}_t(0,0)\ ,\
\rho_{21}(t)={\cal O}^{2,z}_t(0,0)\ ,\
\rho_{22}(t)={\cal O}^{2,-z}_t(0,0)\ .
\eqno(2.11)
$$
Furthermore, since operators $P_i\otimes Q_j$ with $i,j=3,4$ are not
self-adjoint, their expectations can only be measured indirectly,
through the mean values of the projectors
$$
\eqalignno{
&P_{\pm}:=P_{1,2}({\pi\over 4},0)\ ,\
P_{\pm i}:=P_{1,2}({\pi\over 4},-{\pi\over 2})\ ,
&(2.12{\rm a})\cr
&Q_{\pm x}:={|\uparrow_z\rangle\pm|\downarrow_z\rangle\over\sqrt{2}}
{\langle\uparrow_z|\pm\langle\downarrow_z|\over\sqrt{2}}\ ,\
Q_{\pm y}:={|\uparrow_z\rangle\pm i\,|\downarrow_z\rangle\over\sqrt{2}}
{\langle\uparrow_z|\mp i\,\langle\downarrow_z|\over\sqrt{2}}\ .
&(2.12{\rm b})
}
$$

\noindent
Then, from
$$
P_3={P_+ - P_- + i\,P_{+i} - i\,P_{-i}\over2}=P_4^*\ ,\
Q_3={Q_x - Q_{-x} + i\,Q_y - i\,Q_{-y}\over2}=Q_4^*\ ,
\eqno(2.13)
$$
one obtains expressions for all other entries 
$\rho_{i{j\atop k}}(t):={\rm Tr}
\Bigl(\rho_{(2)}(t)P^*_i\otimes Q^*_{j,k}\Bigr)$.
We quote one of them which will be needed in the sequel ( 
the others are reported
in Appendix A), 
$$
\eqalign{
\rho_{43}(t)&={{\cal O}^{1,x}_t({\pi\over 4},0)-
{\cal O}^{1,-x}_t({\pi\over 4},0)\over4}
+i\,{{\cal O}^{1,y}_t({\pi\over 4},0)-
{\cal O}^{1,-y}_t({\pi\over 4},0)\over4}
\cr
&-{{\cal O}^{2,x}_t({\pi\over 4},0)-
{\cal O}^{2,-x}_t({\pi\over 4},0)\over4}
-i\,{{\cal O}^{2,y}_t({\pi\over 4},0)-
{\cal O}^{2,-y}_t({\pi\over 4},0)\over4}
\cr
&+i\,{{\cal O}^{1,x}_t({\pi\over 4},-{\pi\over2})-
{\cal O}^{1,-x}_t({\pi\over 4},-{\pi\over2})\over4}
-\,{{\cal O}^{1,y}_t({\pi\over 4},-{\pi\over2})-
{\cal O}^{1,-y}_t({\pi\over 4},-{\pi\over2})\over4}
\cr
&-i\,{{\cal O}^{2,x}_t({\pi\over 4},-{\pi\over2})-
{\cal O}^{2,-x}_t({\pi\over 4},-{\pi\over2})\over4}
+\,{{\cal O}^{2,y}_t({\pi\over 4},-{\pi\over2})-
{\cal O}^{2,-y}_t({\pi\over 4},-{\pi\over2})\over4}\ .
}\eqno(2.14)
$$

\vskip 1cm

\noindent
{\bf 3. OPEN SYSTEM DYNAMICS INSIDE THE INTERFEROMETER}
\medskip

Neutron interferometry has proved an extremely powerful tool
to investigate gravitational, inertial and 
phase-shifting effects occurring inside the interferometer [23-29].
In the following we will show that neutron interferometry might also be used
to investigate the notion of completely positive open system dynamics.
In order to do that, we consider the case in which neutrons while propagating
inside the interferometric apparatus are 
subjected to weak time-dependent, stochastic magnetic fields coupled to their
spin degree of freedom.

We assume the time-dependent Liouville-Von Neumann evolution equation for the 
$2\times 2$ density matrix $\Sigma$ describing the 
spin degree of freedom to be of the form
$$
\eqalign{
\partial_t \Sigma(t)&=(L_0+L_t)[\Sigma(t)]\ ,\cr
L_0[\Sigma(t)]&:=-i\Bigl[{\omega_0\over 2}\sigma_3\,,\,\Sigma(t)\Bigr]\ ,\quad
L_t[\Sigma(t)]:=-i\Bigl[{\bf V}(t)\cdot\vec{\sigma}\,,\,\Sigma(t)\Bigr]\ ,}
\eqno(3.1)
$$
where $\vec{\sigma}=(\sigma_1,\sigma_2,\sigma_3)$ is the
vector of Pauli matrices, ${\bf V}(t)=(V_1(t),V_2(t),V_3(t))$ 
is proportional to gaussian stochastic magnetic field and
$\displaystyle H_0:={\omega_0\over2}\sigma_3$ 
is due to the coupling to a static magnetic field
along the $z$-direction.
Furthermore, we assume ${\bf V}(t)$ to have  zero mean, 
$\langle{\bf V}(t)\rangle=0$, and 
stationary, real, positive-definite covariance matrix 
${\cal W}(t)=[W_{ij}(t)]$
with entries 
$$
W_{ij}(t-s)=\langle V_i(t)V_j(s)\rangle=W^*_{ij}(t-s)=W_{ji}(s-t)\ .
\eqno(3.2)
$$
Because of the stochastic field $\vec{V}(t)$, the solution
$\Sigma(t)$ of (3.1) is also stochastic;
an effective spin density matrix $\rho(t):=\langle\Sigma(t)\rangle$
is obtained by averaging over the noise.
At time $t=0$ we may suppose spin and noise to
decouple so that  the initial state is
$\rho:=\langle \Sigma(0)\rangle=\Sigma(0)$.
In order to derive an effective time-evolution for $\rho(t)$, 
we follow the so-called convolutionless approach developed in [13].

We average over the noise in the interaction representation, where we set
$$
\eqalignno{
\tilde{\Sigma}(t)&:=\exp{(-tL_0)}[\Sigma(t)]={\rm e}^{-it H_0}
\,\Sigma(t)\,{\rm e}^{it H_0}\ ,
&(3.3{\rm a})\cr
\tilde{\rho}(t)&:=\langle\tilde{\Sigma}(t)\rangle\quad\hbox{and}\quad
\tilde{L}_t:={\rm e}^{-tL_0}\,L_t\,{\rm e}^{tL_0}\ .
&(3.3{\rm b})}
$$
The result is
$$
\tilde{\rho}(t)=
\sum_{k=0}^\infty M_{2k}(t)[\rho]\ ,
\eqno(3.4)
$$
where
$$
M_k(t)[\rho]:=
\int_0^t{\rm d}s_1\int_0^{s_1}{\rm d}s_2
\cdots\int_0^{s_{k-1}}{\rm d}s_k\,
\langle\tilde{L}(s_1)\tilde{L}(s_2)\cdots\tilde{L}(s_k)\rangle[\rho]\ .
\eqno(3.5)
$$
Only even terms
contribute to (3.4) because the stochastic field is assumed to be gaussian.

Denoting by $M_t$ the formal sum in (3.4), a resummation gives
$$
\eqalign{
\partial_t\tilde{\rho}(t)&=\dot{M}_tM_t^{-1}[\tilde{\rho}(t)]\cr
&=\Bigl(\underbrace{\dot{M}_2(t)}_{\hbox{$2$nd order}}\ +\
\underbrace{\dot{M}_4(t)-\dot{M}_2(t)M_2(t)}_{\hbox{$4$th order}}\ +\
\underbrace{\cdots}_{\hbox{higher orders}}\Bigr)[\tilde{\rho}(t)]\ .}
\eqno(3.6)
$$
Since the action of the magnetic field on the travelling neutrons is, by
hypothesis, weak, one can focus on the dominant first term in the expansion, 
neglecting higher order contributions [8-10].
By means of (3.2), the second order contribution can be worked out 
explicitly,
$$
\dot{M}_2(t)[\tilde{\rho}(t)]
=-\sum_{a,b=1}^3\int_0^t{\rm
d}s\,W_{ab}(s)\Bigl[\sigma_a(t),\Bigl[\sigma_b(t-s)\ ,\
\tilde{\rho}(t)\Bigr]\Bigr]\ .
\eqno(3.7)
$$
Returning to the Schr\"odinger representation and using the statistical
independence of the Hamiltonian $H_0$ from the stochastic field, 
it follows that $\rho(t):=\exp(tL_0)[\tilde{\rho}(t)]$ solves
$$
\eqalignno{
&\partial_t\rho(t)=-i\Bigl[H_0\ ,\ \rho(t)\Bigr]\,-\,
\sum_{i,j=1}^3C_{ij}(t)\Bigl[\sigma_i,\Bigl[\sigma_j\ ,\
\rho(t)\Bigr]\Bigr]\ ,
&(3.8{\rm a})\cr
&C_{ij}(t):=\sum_{\ell=1}^3\int_0^t{\rm d}s\, W_{i\ell}(s)
\,U_{\ell j}(-s)\ ,
&(3.8{\rm b})
}
$$
where 
$$
{\cal U}(t):=\pmatrix{\cos\omega_0 t&-\sin\omega_0 t&0\cr
\sin\omega_0 t&\cos\omega_0 t&0\cr
0&0&1}
\eqno(3.9)
$$
is the unitary matrix ${\cal U}(t)=[U_{ij}(t)]$ such that
$\displaystyle
{\rm e}^{-tL_0}[\sigma_i]=\sum_{j=1}^3U_{ij}(t)\sigma_j$.
\bigskip

From (3.2) and (3.9) it follows that
${\cal C}(t):=[C_{ij}(t)]$ is a real matrix and
can thus be decomposed
into symmetric and antisymmetric components.
Correspondingly, the second term on the right hand side of (3.8a) splits into
a commutator with a Hamiltonian and a purely dissipative contribution
$$
\eqalignno{
&\sum_{i,j=1}^3C_{ij}(t)\Bigl[\sigma_i,\Bigl[\sigma_j\ ,\
\rho(t)\Bigr]\Bigr]=i\Bigl[\sum_{i,j,k=1}^3C^A_{ij}(t)\,\epsilon_{ijk}\,
\sigma_k\ ,\ \rho(t)\Bigr]
&(3.10{\rm a})\cr
&\hskip 2cm
+\sum_{i,j=1}^32C^S_{ij}(t)\Bigl({1\over 2}\Bigl\{\sigma_i\sigma_j\ ,\
\rho(t)\Bigr\}\ -\ \sigma_j\,\rho(t)\,\sigma_i\Bigr)\ ,
&(3.10{\rm b})
}
$$
where $C^{A,S}_{ij}(t)=(C_{ij}(t)\mp C_{ji}(t))/2$.

When the coupling between system and stochastic field is weak, the memory
effects in (3.8a) should not be physically relevant; therefore, the use of a
Markov approximation is in general justified.%
\footnote{$^\dagger$}{More precisely, one can show that a linear, local in time
subdynamics is the result of a limiting procedure in which the coupling
constant $g$ between system and external environment, and the ratio
$\tau/T$ between the typical time scale of the system and the decay time
of the correlations in the environment, become small.[8-10]
The quantities $g$ and $\tau/T$ regulate both the expansion (3.6)
and the Markovian approximation of (3.8b).}
In practice this is done by extending to $+\infty$ the 
upper limit of the integral in (3.8b); the resulting equation of motion has no
explicit time-dependence,
$$
\eqalignno{
\partial_t\rho(t)&=-i\Bigl[H\ ,\ \rho(t)\Bigr]\ +\ L_{D}[\rho(t)]\ ,
&(3.11{\rm a})
\cr
H&=H_0+H_D\ ,\qquad
H_D:=\sum_{i,j,k=1}^3C^A_{ij}\epsilon_{ijk}\,\sigma_k\ ,
&(3.11{\rm b})\cr
L_{D}[\rho(t)]&:=\sum_{i,j=1}^3L^{D}_{ij}\Bigl(-{1\over 2
}\Bigl\{\sigma_i\sigma_j\ ,\
\rho(t)\Bigr\}\ +\ \sigma_j\,\rho(t)\,\sigma_i\Bigr)\ ,
&(3.11{\rm c})
}
$$
where
$C^A_{ij}$ and $L^{D}_{ij}$ are the entries
of the real matrices  
$$
\eqalignno{
{\cal C}_A&:=\int_0^{+\infty}{\rm d}s\, {{\cal W}(s)\,{\cal U}(-s)\,-
\,{\cal U}(s)\,{\cal W}(-s)\over 2}\ ,&(3.12{\rm a})\cr
{\cal L}_{D}&:=\int_0^{+\infty}{\rm d}s\, \Bigl({\cal W}(s)\,{\cal U}(-s)\,+
\,{\cal U}(s)\,{\cal W}(-s)\Bigr)\ .&(3.12{\rm b})
}
$$

The Hamiltonian contribution is skew-symmetric, while
the purely dissipative one, $L_{D}[\cdot]$, is symmetric; this latter  
makes the time-evolution irreversible, but preserves probability because 
${\rm Tr}\Bigl(L_{D}[\rho(t)]\Bigr)=0$.
The solutions of (3.11a) thus constitute a semigroup of linear maps 
$\Gamma_{t}: \rho\longmapsto\rho(t)
:=\Gamma_{t}[\rho]$
such that
$\Gamma_{t+s}=\Gamma_{t}\circ\Gamma_{s}$, $s,t\geq 0$
and ${\rm Tr}(\Gamma_{t}[\rho])={\rm Tr}\rho$.
It remains to be checked whether they preserve the positivity
of spin density matrices, that is whether the $\Gamma_t$'s are, in short, 
positive maps on the spin states, which is the first request for physical
consistency.

There is, however, a further constraint that has to be respected for physical
consistency.
Indeed, if the neutron interferometer is placed in a stochastic classical 
magnetic field of the kind described above, 
the translational degree of freedom is not affected
and the effective state
$\rho_{(2)}(t)$ at the exit from the interferometer will be
$\rho_{(2)}(t)=({\bf I}_{2}\otimes\Gamma_{t})[\rho_{(2)}]$, 
where ${\bf I}_{2}$ denotes the 
identity operation on the first factors in (2.4).
It turns out that the positivity of the maps $\Gamma_t$ does not guarantee
the positivity and thus the physical consistency of the maps 
${\bf I}_{2}\otimes\Gamma_{t}$; for this the stronger notion of complete
positivity has to be imposed on the maps $\Gamma_t$.

We shall later investigate these notions in more technical detail,
for the moment we observe that generating fluctuating magnetic backgrounds
with certain 
decaying properties of their covariance matrix (3.2), 
one may have experimental
access to some physical situations of theoretical interest that we will
present below.

\bigskip

\noindent
{\bf 3.1 White noise}
\smallskip

\noindent
The stochastic magnetic field has
white-noise correlations
$$
W_{{ij}}(t-s)=\langle V_i(t)V_j(s)\rangle=W_{ij}\,\delta(t-s)\ ,
\eqno(3.13)
$$
where ${\cal W}:=[W_{ij}]$ is time-independent, symmetric and
positive-definite.
Then,
${\cal C}_{A}=0$ and ${\cal L}_{D}={\cal W}$; furthermore,  
writing ${\cal W}={\cal A}^{2}$ with ${\cal A}=[a_{ij}]$
real and symmetric, the 
dissipative term in (3.11a) reads
$$
L_D[\rho]=\sum_{k}A_k\,\rho\,A_k\,-\,{1\over 2}\,
\Bigl\{\sum_kA_k^2\,,\,\rho\Bigr\}\ ,
\eqno(3.14)
$$
with self-adjoint $A_k:=\sum_{i=1}^3a_{ki}\sigma_{i}$.
This is a particular instance of Lindblad's theorem [30,31] which states that
a family of linear transformations $\Gamma_{t}:\rho\mapsto\rho(t)$ on the 
$D$-dimensional density matrices is a quantum dynamical semigroup of 
probability-preserving, completely positive maps if and only if it is
generated by
the equation of motion
$$
\partial_t\rho(t)=-i\Bigl[H\,,\,\rho(t)\Bigr]\,-\,
{1\over2}\Bigl\{\sum_\ell A^*_\ell A_\ell\,,\,\rho(t)\Bigr\}\,+\,
\sum_\ell A_\ell\rho(t)A^*_\ell\ ,
\eqno(3.15)
$$
where the $A_\ell$'s are $D$-dimensional matrices with adjoint $A^*_\ell$
such that the series are (norm-) convergent.
On the other hand, if $L_D$ in (3.11a) is as in (3.15), then the corresponding
${\cal L}_D$ is positive definite.
Further, note that, because of (3.13), all higher order terms in the expansion
(3.6) identically vanish, so that the evolution equation (3.15) is in this case
exact [22].

\bigskip

\noindent
{\bf 3.2 Diagonal covariance matrix}
\smallskip

\noindent
The stochastic field has no off-diagonal 
correlations, while
$$
\langle V_{1}(t)V_{1}(s)\rangle=\langle V_{2}(t)V_{2}(s)\rangle
=g^{2}B_{1}^{2}{\rm e}^{-\lambda|t-s|}\ ,\quad
\langle V_{3}(t)V_{3}(s)\rangle=g^{2}B_{3}^{2}{\rm e}^{-\mu|t-s|}\ ,
\eqno(3.16)
$$
where $B_i$ are constant magnetic field intensities and
$g$ is proportional to the neutron magnetic moment.
Then,
$$
{\cal C}_{A}={g^{2}\omega_{0}B_{1}^{2}\over \lambda^2+\omega_0^{2}}\pmatrix{
0&1&0\cr
-1&0&0\cr0&0&0}\ ,\quad
{\cal L}_{D}=2g^{2}\pmatrix{
{\lambda B_{1}^2\over\lambda^{2}+\omega_{0}^{2}}&0&0\cr
0&{\lambda B_{1}^2\over\lambda^{2}+\omega_{0}^{2}}&0\cr
0&0&{B_{3}^2\over\mu}}\ .\eqno(3.17)
$$
Setting
$$\Delta\omega:={4g^{2}B_{1}^{2}\omega_{0}\over
\lambda^{2}+\omega_{0}^{2}}\  ,\quad
\gamma:={4g^2B_1^2\lambda\over\lambda^2+\omega_0^2}\ ,\quad
a:={B_3^2\over\mu}+{2g^{2}B_{1}^{2}\omega_{0}\over
\lambda^{2}+\omega_{0}^{2}}\ ,
\eqno(3.18)
$$
the matrix in (3.12b) becomes
$$
{\cal L}_{D}={1\over 2}\pmatrix{
\gamma&0&0\cr
0&\gamma&0\cr
0&0&2a-\gamma}\ .
\eqno(3.19)
$$
The reason for such a parametrization will become clear in the next 
section.
\hfill\break
Given ${\cal L}_D$, the entries of the spin matrix
$\rho(t)=\pmatrix{\rho_1(t)&\rho_3(t)\cr\rho_4(t)&\rho_2(t)}$are 
readily showed to satisfy
the Bloch-Redfield equations [32]
$$
\dot{\rho}_1=-\gamma\rho_1+\gamma\rho_{2}\ ,\quad
\dot{\rho}_3=-i(\omega_{0}+\Delta\omega)\rho_3-2a\rho_3
\eqno(3.20)
$$
and $\dot{\rho}_2=-\dot{\rho}_1$, $\dot{\rho}_4=(\dot{\rho}_3)^*$.
The coefficients $2\gamma$ and $2a$ are the inverse of the relaxation times
$T_{1}$ 
and $T_{2}$ of the diagonal, respectively off-diagonal elements of 
$\rho(t)$; from the positivity condition $2a-\gamma\geq0$, it follows that: 
$1/T_{2}\geq 1/2T_{1}$.
In [13,14] it is showed that this typical order relation
can be reversed by setting $B_{3}=0$ and 
keeping $4$-th order terms in (3.6).
In such a case, however,
$1/T_{2}< 1/2T_{1}$ implies $a-\gamma/2<0$ and ${\cal L}_D$ in (3.19)
is no longer positive-definite.
\hfill\break
By Lindblad's theorem, the argument of case 1. implies that
the corresponding dynamical maps
$\Gamma_t$ generated through (3.11a) can not be completely positive.
We shall see in the next section that at least they preserve positivity.

\bigskip

\noindent
{\bf 3.3 Single component field correlation}
\smallskip

\noindent
The stochastic magnetic field is along the $x$-direction,
$\vec{V}(t)=(V_1(t),0,0)$, with
$$
\langle V_{1}(t)V_1(s)\rangle=g^2B^2\,{\rm e}^{-\lambda|t-s|}\ .
\eqno(3.21)
$$
Then, 
$$
{\cal C}_{A}={g^{2}\omega_{0}B^{2}\over 2(\lambda^2+\omega_0^{2})}\pmatrix{
0&1&0\cr
-1&0&0\cr0&0&0}\ ,\quad
{\cal L}_{D}={g^{2}B^{2}\over\lambda^{2}+\omega_0^{2}}
\pmatrix{
2\lambda&\omega_{0}&0\cr
\omega_{0}&0&0\cr
0&0&0}\ .\eqno(3.22)
$$
Unless $\omega_0=0$ the matrix ${\cal L}_{D}$ is not positive-definite.
By analogy with the parametrization of the previous example, we set
$$\Delta\omega:={2g^{2}B^{2}\omega_{0}\over
\lambda^{2}+\omega_{0}^{2}}\  ,\quad
\gamma:={2g^2B^2\lambda\over\lambda^2+\omega_0^2}\ ,\quad
b:=-{\Delta\omega\over 2}\ .
\eqno(3.23)
$$
Then, the corresponding Bloch-Redfield equations for the entries of 
$\rho(t)$ read
$$
\dot{\rho}_1=-\gamma\rho_1+\gamma\rho_{2}\ ,\quad
\dot{\rho}_3=-i(\omega_{0}+\Delta\omega)\rho_3-\gamma\rho_3
+\gamma\rho_{4}+2ib\rho_{4}\ ,
\eqno(3.24)
$$
and $\dot{\rho}_2=-\dot{\rho}_1$, $\dot{\rho}_4=(\dot{\rho}_3)^*$. 


\vskip 1cm

\noindent
{\bf 4.  COMPLETE POSITIVITY VS SIMPLE POSITIVITY }
\medskip
 
As already remarked, physical consistency demands
that $\Gamma_{t}$ preserve the positivity of initial density matrices
$\rho$ describing the neutron spin degree of freedom.

In order to check whether this is so in the preceeding cases,
it is convenient to decompose the spin density matrices 
$\rho$
by means of the Pauli matrices $\sigma_{j}$, $j=1,2,3$ plus the 
$2$-dimensional identity matrix $\sigma_{0}$, namely
$\rho=\sum_{\mu=0}^3\rho^\mu\sigma_\mu$.
In such a way, density matrices can be represented
as $4$-dimensional ket-vectors $|\rho\rangle=(\rho^0,\rho^1,\rho^2,\rho^3)$,
where
$$
\rho^{0}={\rho_{1}+\rho_{2}\over2}\ ,\
\rho^{1}={\rho_{3}+\rho_{4}\over2}\ ,\
\rho^{2}={\rho_{4}-\rho_{3}\over2i}\ ,\
\rho^{3}={\rho_{1}-\rho_{2}\over2}\ .
\eqno(4.1)
$$

Since they operate linearly,
the commutator and the purely dissipative term $L_D[\ \cdot\ ]$
in
(3.11a) act on the vectors $|\rho\rangle$ as
a skew-symmetric matrix 
$$
{\cal H}=-2\pmatrix{0&0&0&0\cr
0&0&h^{3}&-h^{2}\cr
0&-h^{3}&0&h^{1}\cr
0&h^{2}&-h^{1}&0}\ ,\quad h^{i}\in{\bf R}\ ,
\eqno(4.2)
$$
respectively as a real, symmetric matrix
$$
{\cal D}=-2\pmatrix{0&0&0&0\cr
0&a&b&c\cr
0&b&\alpha&\beta\cr
0&c&\beta&\gamma}\ .
\eqno(4.3)
$$
The connection with ${\cal L}_{D}$ in (3.12b) is readily derived to be
$$
{\cal L}_{D}={1\over 2}\pmatrix{
\alpha+\gamma-a&-2b&-2c\cr
-2b&a+\gamma-\alpha&-2\beta\cr
-2c&-2\beta&a+\alpha-\gamma}
\ ,
\eqno(4.4)
$$
which explains the parametrization used in the previous section.

In this representation, the time-evolution equation (3.11a) reads
$$
\partial_{t}|\rho(t)\rangle=({\cal H}+{\cal D})|\rho(t)\rangle\ .
\eqno(4.5)
$$
In order to find necessary and sufficient conditions for $\Gamma_t$ to be
positivity preserving, we now proceed in a few steps.

First, since $\dot{\rho}^0=0$, the trace is conserved, thus 
$\rho^{0}(t)=1/2$; therefore, 
the positivity of $\Gamma_{t}[\rho]$ is ensured if
${\rm Det}[\rho(t)]=1/4-\sum_{j=1}^{3}(\rho^{j})^{2}\geq 0$.

Second, the time-derivative of the determinant at $t=0$ must be positive
whenever $\sum_{j=1}^3(\rho^j)^2=1/4$ (so that ${\rm Det}[\rho]=0$),
otherwise one eigenvalue would become negative for $t>0$.
Using (4.3,4.5), we  thus get the necessary condition
$$
{{\rm d Det}[\rho(0)]\over {\rm d}t}=-2\sum_{i,j=1}^{3}
{\cal D}_{ij}\rho^{i}\rho^{j}\geq0\ .
\eqno(4.6)
$$
By varying  $\rho^{j}$, while keeping
$\sum_{j}(\rho^{j})^{2}=1/4$, it follows that 
$$
{\cal D}^{(3)}:=-2\pmatrix{a&b&c\cr
b&\alpha&\beta\cr
c&\beta&\gamma}
\eqno(4.7)
$$ 
must be negative definite which in turn implies
$$
a\geq0\ ,\quad
a\alpha\geq b^{2}\ ,\quad{\rm Det}{\cal D}^{(3)}\geq0
\ .
\eqno(4.8)
$$ 

Third, conditions (4.8) are also sufficient for $\Gamma_t$ to preserve 
positivity.
In fact, since $-{\cal D}\geq0$,  
we can write $-{\cal D}={\cal B}^2$ with $\cal B$ symmetric.
Then, the term in the right hand side of the equality in (4.6) is 
given by $\|{\cal B}|\rho\rangle\|^2$.
Let us suppose ${\rm Det}[\rho(t')]<0$, at time $t'>0$; 
it follows that 
${\rm Det}[\rho(t^*)]=0$ at some time $t^*$ such that
$0\leq t^*<t'$.
Thus, ${\cal B}|\rho(t^*)\rangle=0$, otherwise
${\rm Det}[\rho(t)]>0$ for $t\geq t^*$; but this implies
$|\rho(t)\rangle=|\rho(t^*)\rangle$ for
all $t\geq t^*$ under the time-evolution
$$
|\rho\rangle\longmapsto
|\rho(t)\rangle={\rm e}^{t{\cal D}}|\rho\rangle=\sum_{\ell=0}
{t^\ell\over\ell !}{\cal B}^{2\ell}|\rho\rangle\ .
\eqno(4.9)
$$
Therefore, as well as the standard dynamics generated by a Hamiltonian operator,
the dynamics (4.9) is positivity-preserving;
via the Lie-Trotter product formula, it then follows that
the time-evolution generated by $\cal H+D$,
$$
{\cal G}_t:=\exp{\Bigl(t({\cal H}+{\cal D})\Bigr)}=\lim_{n\to\infty}
\Bigl({\rm e}^{t/n{\cal H}}{\rm e}^{t/n{\cal D}}\Bigr)^n\ ,
\eqno(4.10)
$$
preserves positivity, too.

However, even if $\Gamma_t$ preserves positivity of the states describing the
neutron spin degree of freedom,
this does not guarantee that
${\bf I}_{2}\otimes\Gamma_{t}$ preserves the positivity of
states in which the spin is entangled with another degree of freedom. 
For neutrons in the interferometric apparatus of the previous
section, such a request is crucial since the maps
${\bf I}_{2}\otimes\Gamma_{t}$ tell us how do evolve in time
states $\rho_{(2)}$ describing both
the magnetic and translational degrees of freedom.
A theorem of Choi [12] states ${\bf I}_2\otimes\Gamma_t$
to be positivity preserving if and only if
$\Gamma_t$ is completely positive.

Among the neutron states $\rho_{(2)}$ propagating through
the interferometer, those without correlations between
translational and spin degrees of freedom are of the form $\rho_{{\rm space}}
\otimes\rho_{{\rm spin}}$ or are linear mixtures of them.
If the $\Gamma_t$ are positivity preserving these tensor-product states
remain positive in the course of time;
indeed,
$$
0\leq\rho_{{\rm space}}
\otimes\rho_{{\rm spin}}\longmapsto
{\rm I}_{2}\otimes\Gamma_{t}[\rho_{{\rm space}}
\otimes\rho_{{\rm spin}}]=\rho_{{\rm space}}
\otimes\Gamma_{t}[\rho_{{\rm spin}}]\geq0\ .
\eqno(4.11)
$$
However, it is not so for entangled states.

Let us take $p=-q=1/\sqrt{2}$ in (2.2),
so that the initial beam state is antisymmetric in
the two degrees of freedom.
If we ask ${\bf I}_{2}\otimes\Gamma_{t}$ to preserve the positivity of
$\rho_\Psi$, it must hold that
$$
\Delta_{\Phi}(t):= 
\langle\Phi|({\bf I}_{2}\otimes\Gamma_{t})\,[\rho_{\Psi}]|\Phi\rangle\, \geq0
\eqno(4.12)
$$
for all $\Phi$.
If $\Phi$ is orthogonal to $\Psi$,
the fact that $\Delta_{\Phi}(0)=0$ implies
$$
\left.{{\rm d}\over{\rm d}t}\Delta(t)\right|_{t=0}=
\langle\Phi|
({\bf I}_{2}\otimes L_{D})\,[\rho_{\Psi}]|\Phi\rangle\geq0\ .
\eqno(4.13)
$$
By varying $\Phi$ in the $3$-dimensional subspace orthogonal to $\Psi$ we
obtain the inequalities 
$$
\eqalign{
&2R\equiv\alpha+\gamma-a\geq0\ ,\quad RS\geq b^2\cr
&2S\equiv a+\gamma-\alpha\geq0\ ,\quad RT\geq c^2\cr
&2T\equiv a+\alpha-\gamma\geq0\ ,\quad ST\geq\beta^2\cr
&RST\geq 2\, bc\beta+R\beta^2+S c^2+T b^2\ .}                
\eqno(4.14)
$$
These inequalities are stronger than the ones in (4.8) and
must necessarily be satisfied if we want to
avoid that the maps ${\bf I}_{2}\otimes\Gamma_{t}$
become physically inconsistent by generating negative probabilities
out of initially entangled beam states.

Furthermore, using (4.4), inequalities (4.14) amount to 
the positivity of the matrix ${\cal L}_{D}=[L^{D}_{ij}]$ 
of the coefficients of the dissipative term in (3.11c) and thus
they imply the complete positivity of 
the time-evolution $\Gamma_t$.

Let us now discuss the three cases introduced in Section 3.

\bigskip

\noindent
{\bf 4.1 White noise}
\smallskip

\noindent
As already noticed, the evolution equation (3.11) is now exact. The matrix
${\cal D}^{(3)}$ takes the general form (4.7), so that provided the inequalities
(4.14) are satisfied, the integrated time evolution $\Gamma_t$ results
completely positive.

\bigskip

\noindent
{\bf 4.2 Diagonal covariance matrix}
\smallskip

\noindent
In this case one finds:
$$
{\cal D}^{(3)}=-2\pmatrix{a&0&0\cr
0&a&0\cr
0&0&\gamma}\ .
\eqno(4.15)
$$
The corresponding dynamics is completely positive only
when $1/T_{2}\geq 1/2T_{1}$ ($a\geq\gamma/2$); it is positive, but not completely 
positive, when $1/T_{2}<1/2T_{1}$ ($a<\gamma/2$). 

Analytic solutions of the equation of motion corresponding to 
(4.15) are readily calculated; in vectorial representation one has 
$$
\eqalign{
\rho^{0}(t)&=\rho^0\cr
\rho^{1}(t)&={\rm e}^{-2at}\Bigl\{\rho^{1}\cos\omega t\,-\,
\rho^{2}\sin\omega t\Bigr\}\cr
\rho^{2}(t)&={\rm e}^{-2at}\Bigl\{\rho^{1}\sin\omega t\,+\,
\rho^{2}\cos\omega t\Bigr\}\cr
\rho^{3}(t)&={\rm e}^{-2\gamma t}\rho^{3}}\ ,
\eqno(4.16)
$$
with $\omega:=\omega_0+\Delta\omega$. The difference between complete positivity and 
positivity shows up in different order relations between the decay
diagonal and off-diagonal relaxation, 
that is either $a\geq\gamma/2$ or $a<\gamma/2$.

However, the true physical meaning of complete positivity of $\Gamma_t$
becomes evident when the state $\rho_\Psi$ in (2.2), with $p=-q=1/\sqrt{2}$, 
evolves in time according to
${\bf I}_{2}\otimes\Gamma_{t}$.
In vectorial representation
$|Q_1\rangle=1/2(1,0,0,1)$, $|Q_2\rangle=1/2(1,0,0,-1)$,
$|Q_3\rangle=1/2(0,1,i,0)$ and $|Q_4\rangle=1/2(0,1,-i,0)$,
thus, using (4.16) one obtains
$$
\rho_{(2)}(t)=({\bf I}_{2}\otimes\Gamma_t)\,[\rho_\Psi]=
{1\over 2}\Bigl(
P_{1}\otimes Q_{2}(t)\,+\,P_{2}\otimes Q_1(t)\,-\,
P_{3}\otimes Q_{4}(t)\,-\,P_{4}\otimes Q_3(t)\Bigr)\ ,
\eqno(4.17)
$$
\noindent
with
$$
\eqalign{
Q_{1}(t)&={1\over 2}\pmatrix{
1+{\rm e}^{-2\gamma t}&0\cr
0&1-{\rm e}^{-2\gamma t}}\  ,\quad
Q_{3}(t)={\rm e}^{-t(2a+i\omega)}\pmatrix{0&1\cr0&0}
\cr
\cr
Q_{2}(t)&={1\over 2}\pmatrix{
1-{\rm e}^{-2\gamma t}&0\cr
0&1+{\rm e}^{-2\gamma t}}\ ,\quad
Q_{4}(t)={\rm e}^{-t(2a-i\omega)}\pmatrix{0&0\cr1&0}
\ .}
\eqno(4.18)
$$
\noindent
Therefore,
$$
\rho_{(2)}(t)=\pmatrix{
E_{-}(t)&0&0&0\cr
0&E_{+}(t)&F(t)&0\cr
0&F^{*}(t)&E_{+}(t)&0\cr
0&0&0&E_{-}(t)},\quad 
E_{\pm}:={1\pm{\rm e}^{-2\gamma t}\over 4},\quad
F:=-{{\rm e}^{-t(2a-i\omega)}\over 2}\ .
\eqno(4.19)
$$
The eigenvalues of the state $\rho_{(2)}(t)$ at the exit of
the interferometer are 
$\lambda_{1,2}(t)=E_{-}(t)$ and 
$$
\lambda_{\pm}(t)={1+{\rm e}^{-2\gamma t}\pm2{\rm e}^{-2 a t}\over 4}\ .
\eqno(4.20)
$$
Let $a<\gamma/2$, that is
let $\Gamma_t$ to be positive, but not completely positive;
then, since $\lambda_{-}(0)=0$ and 
${\rm d}\lambda_{+}(0)/{\rm d}t=(2a-\gamma)/2$, there is
a whole range of $t$ where $\lambda_{+}(t)<0$ and $\rho_{(2)}(t)$ loses
physical meaning.
On the other hand, if $a\geq\gamma/2$, $\lambda_{+}(t)\geq 0$, for 
all $t$.

\bigskip

\noindent
{\bf 4.3 Single component field correlation}
\smallskip

\noindent
The matrix ${\cal D}^{(3)}$ in (4.7) has now also off-diagonal terms:
$$
{\cal D}^{(3)}=-2\pmatrix{0&b&0\cr
b&\gamma&0\cr
0&0&\gamma}\ ,\quad
b=-{\Delta\omega\over2}\ ;
\eqno(4.21)
$$
only when these are zero, {\it i.e.} $\Delta\omega=0$, positivity is preserved.
This can be seen also by considering the integrated time evolution;
in the vectorial representation, one explicitly finds:
$$
\eqalign{
\rho^{0}(t)&=\rho^0\cr
\rho^{1}(t)&={\rm e}^{-\gamma t}\Bigl\{\rho^{1}\Bigl(\cosh\delta t\,+\,
{\gamma\over\delta}\sinh\delta t\Bigr)\,-\,\rho^{2}\
{\omega+2b\over\delta}\sinh\delta t\Bigr\}\cr
\rho^{2}(t)&={\rm e}^{-\gamma t}
\Bigl\{\rho^{1}\ {\omega-2b\over\delta}\sinh\delta t\,+\,
\rho^{2}\Bigl(\cosh\delta t\,-\,{\gamma\over\delta}\sinh\delta t\Bigr)
\Bigr\}\cr
\rho^{3}(t)&={\rm e}^{-2\gamma t}\rho^{3}}\,
\eqno(4.22)
$$
with $\delta:=\sqrt{\gamma^{2}+(4b^2-\omega^{2})}$. One notes that, 
besides developing negative eigenvalues because of lack of positivity, 
this evolution cause $\rho^{(1,2)}(t)$ to 
diverge with $t\mapsto+\infty$, when $\delta>\gamma$.

\bigskip

Although apparently formal, these results are far from being academic:
indeed, as already stressed at the end of Section 2, the entries of
$\rho_{(2)}$ are directly accessible to the experiment.%
\footnote{$^\dagger$}{For instance, in the case {\bf 4.2} above,
it follows from (2.11) that
$E_-(t)={\cal O}^{1,z}_t(0,0)$ and $E_+(t)={\cal O}^{2,z}_t(0,0)$,
while $F(t)$ coincides with the expression in (2.14).}
By modulating a background magnetic field close to the stochastic
properties investigated in the previous three cases, one might reproduce
experimentally the conditions for three different reduced dynamics
and check their consequences.

Then, one may conclude that reduced, markovian time-evolutions 
$\Gamma_t$ must be not only positive, but also completely positive,
since lack of any of these constraints results
in experimentally detectable inconsistencies.

Clearly, the use of one reduced dynamics instead of another depends 
on the markovian approximation used to derive it and whether, given the
properties of the stochastic field, it was justified or not.
It thus seems appropriate to conclude that, 
whenever a semigroup composition law is
expected, the physically appropriate markovian approximations are those which
lead to reduced dynamics consisting of completely positive maps 
$\Gamma_t$ [17].

\vskip 1cm

\noindent
{\bf 5. NONCONTEXTUALITY AND DISSIPATION}
\medskip

We now examine to what extent the Clauser-Horne-Shimony-Holt inequalities 
(2.10) are modified by the presence of a stochastic magnetic field with 
covariance matrix as in the cases studied in Section 3. 
We first notice that the mean values (2.9) can be written
$$
C_t(\vartheta,\varphi;{\bf n})=2\Bigl(
{\cal O}^{1,{\bf n}}_t(\vartheta,\varphi)-
{\cal O}^{1,-{\bf n}}_t(\vartheta,\varphi)\Bigr)\,
-\,{\rm Tr}_2\Bigl(\rho_{{\rm spin}}(t)B({\bf n})\Bigr)\ ,
\eqno(5.1)
$$
where ${\rm Tr}_{2}$ denotes the trace over the spin
degree of freedom and
$\rho_{{\rm spin}}(t)={\rm Tr}_1\rho_{(2)}(t)$, ${\rm Tr}_1$ 
denoting the trace over the translational degree of freedom.

For sake of simplicity, we again consider an initial beam state with
$p=-q=1/\sqrt{2}$; then,
$$
\eqalign{
C_t(\vartheta,\varphi;{\bf n})&=2\Bigl(
{\cal O}^{1,{\bf n}}_t(\vartheta,\varphi)-
{\cal O}^{1,-{\bf n}}_t(\vartheta,\varphi)\Bigr)\cr
&=\sin^2\vartheta\ {\rm Tr}_2\Bigl(Q_2(t)B({\bf n})\Bigr)+\cos^2\vartheta\
{\rm Tr}_2\Bigl(Q_1(t)B({\bf n})\Bigr)\cr
&\hskip 2cm
-\,
\sin2\vartheta\
{\cal R}e\Biggl({\rm e}^{-i\varphi}{\rm Tr}_2\Bigl(
Q_4(t)B({\bf n})\Bigr)\Biggr)\ .}
\eqno(5.2)
$$
Further, we shall take $h^1=h^2=0$ and $h^3=\omega_0/2$ in the hamiltonian part
(4.2) of the evolution equation (4.5).

From (4.10) it follows that, in vectorial representation,
the time-evolution operator ${\cal G}_t$ acts
on initial states $|\rho\rangle$ as 
$$
{\cal G}_t=\pmatrix{1&0&0&0\cr
0&{\cal G}_{11}(t) &{\cal G}_{12}(t)&{\cal G}_{13}(t)\cr
0&{\cal G}_{21}(t)&{\cal G}_{22}(t)&{\cal G}_{23}(t)\cr
0&{\cal G}_{31}(t)&{\cal G}_{32}(t)&{\cal G}_{33}(t)}\ .
\eqno(5.3)
$$
Also, the mean value of an observable $X=\sum_{\nu=0}^3X^\nu\sigma_\nu$
with respect to $\rho$ is given by 
${\rm Tr}(X\rho)=2\sum_{\nu=0}^3X^\nu\rho^\nu$.
Further, the observable $B({\bf n})=Q_{{\bf n}}-Q_{-{\bf n}}$ relative to 
${\bf n}=(n_1,n_2,n_3)$ corresponds to the vector
$|B({\bf n})\rangle=(0,n_1,n_2,n_3)$; thus one computes 
$$
{\rm Tr}_2\Bigl(Q_{1,2}(t)B({\bf n})\Bigr)=\pm{\bf G}(t)\cdot{\bf n}\ ,\qquad
{\rm Tr}_2\Bigl(Q_4(t)B({\bf n})\Bigr)={\bf F}(t)\cdot{\bf n}\ ,
\eqno(5.4)
$$
where
$$
\eqalignno{
{\bf G}(t)&=\Bigl({\cal G}_{13}(t),\
{\cal G}_{23}(t),\ {\cal G}_{33}(t)\Bigr)
\hskip 1cm
{\rm and}&
(5.5{\rm a})\cr
{\bf F}(t)&=\Bigl({\cal G}_{11}(t)-i{\cal G}_{12}(t),\
{\cal G}_{21}(t)-i{\cal G}_{22}(t),\
{\cal G}_{31}(t)-i{\cal G}_{32}(t)\Bigr)\ .
&(5.5{\rm b})}
$$
Finally, the mean values (5.2) read
$$
C_t(\vartheta,\varphi;{\bf n})={\bf n}\cdot\Bigl[
\cos2\vartheta\,{\bf G}(t)\,-\,\sin2\vartheta\,
{\cal R}e\Bigl({\rm e}^{-i\varphi}{\bf F}(t)\Bigr)\Bigr]\ .
\eqno(5.6)
$$
We shall now discuss the explicit behaviour of $C_t(\vartheta,\varphi;{\bf n})$
in the three cases introduced in Section 3, and further analyzed in the
previous section.

\bigskip

\noindent
{\bf 5.1 White noise}
\smallskip

\noindent
Though analytic expressions of ${\cal G}_t$ are obtainable in the general
case of a stochastic magnetic field with white noise correlations, 
these are rather involved and scarcely illuminating.
More conveniently, one may suppose the dissipation $\cal D$ in (4.3) to be small
in comparison to the Hamiltonian contribution due to $H_0$.
In practice, we assume the parameters $a$, $b$, $c$, $\alpha$, $\beta$ 
and $\gamma$ small with respect to
$\omega=\omega_0+\delta\omega\simeq\omega_0$
and proceed with a perturbative expansion (for more details compare
[18,19,21]).
To first order in the
parameters the 
entries of ${\cal G}_t$ are as reported in the Appendix B.
Using them, one calculates
$$
\eqalignno{
&G_1(t)=
-{4|C|\over\omega_0}\sin{\omega_0 t\over 2}\cos\Big({\omega_0 t\over2}+\phi_C\Big)\ ,
&(5.7{\rm a})\cr
&G_2(t)=
{4|C|\over\omega_0}\sin{\omega_0 t\over 2}\sin\Big({\omega_0 t\over2}+\phi_C\Big)\ ,
&(5.7{\rm b})\cr
&G_3(t)={\rm e}^{-2\gamma t}\ , &(5.7{\rm c})\cr
}
$$
where
$|C|^2=c^2+\beta^2$ and $\tan\phi_C=\beta/c$,
$$
\eqalignno{
{\cal R}e\Bigl({\rm e}^{-i\varphi}F_1(t)\Bigr)&=
{\rm e}^{-(a+\alpha)t}\cos(\omega_0t-\varphi)\,+\,
{|B|\over\omega_0}\sin\omega_0t\,
\cos(\varphi+\phi_B)\ ,
&(5.8{\rm a})\cr
{\cal R}e\Bigl({\rm e}^{-i\varphi}F_2(t)\Bigr)&=
{\rm e}^{-(a+\alpha)t}\sin(\omega_0t-\varphi)\,-\,
{|B|\over\omega_0}\sin\omega_0t\,
\sin(\varphi-\phi_B)\ ,
&(5.8{\rm b})\cr
{\cal R}e\Bigl({\rm e}^{-i\varphi}F_3(t)\Bigr)&=
-{4|C|\over\omega_0}\sin{\omega_0\over 2}t
\cos\Big({\omega_0\over 2}t-\varphi-\phi_C\Big)\ ,
&(5.8{\rm c})}
$$
where $|B|^2=(a-\alpha)^2+4b^2$ and $\tan\phi_B=2b/(\alpha-a)$.
It thus follows that
$$
\eqalign{
C_t(\vartheta,\varphi;{\bf n})&=n_1\Bigl[-{4|C|\over\omega_0}
\cos2\vartheta\sin{\omega_0\over2}t\cos\Big({\omega_0\over2}t+\phi_C\Big)\Bigr.\cr
&\qquad\Bigl.
-\sin2\vartheta\,\Bigl(
{\rm e}^{-t(a+\alpha)}\cos(\omega_0t-\varphi)+{|B|\over\omega_0}
\sin\omega_0t\cos(\varphi+\phi_B)\Bigl)\Bigr]\cr
&+n_2\Bigl[
{4|C|\over\omega_0}\cos2\vartheta\sin{\omega_0\over2}t\sin({\omega_0\over2}t
+\phi_C)\Bigl.\cr
&\qquad\Bigl.
-\sin2\vartheta\Bigl({\rm e}^{-t(a+\alpha)}\sin(\omega_0t-\varphi)-
{|B|\over\omega_0}\sin\omega_0t\sin(\varphi-\phi_B)\Bigr)\Bigr]\cr
&+n_3\Bigl[
{\rm e}^{-2\gamma t}\cos2\vartheta\,-\,
{4|C|\over\omega_0}\sin2\vartheta\sin{\omega_0\over2}t\,
\cos\Big({\omega_0\over2}t-\varphi-\phi_C\Big)\Bigr]\ .}
\eqno(5.9)
$$

\bigskip

\noindent
{\bf 5.2 Diagonal covariance matrix}
\smallskip

\noindent
For the stochastic magnetic fields with correlation matrices
as in (3.16) we have
$$
{\bf G}(t)=\Bigl(0,\ 0,\ {\rm e}^{-2\gamma t}\Bigr)\ ,\quad
{\bf F}(t)={\rm e}^{-t(2a-i\omega)}\,(1,\ -i,\ 0)\ .
\eqno(5.10)
$$
With these expressions one easily derives
$$
C_t(\vartheta,\varphi;{\bf n})=
{\rm e}^{-2\gamma t}\cos2\vartheta\, n_3
-{\rm e}^{-2at}\sin2\vartheta\,\Bigl(n_1\cos(\omega t-\varphi)+
n_2\sin(\omega t-\varphi)\Bigr)\ .
\eqno(5.11)
$$

\vfill\eject

\noindent
{\bf 5.3 Single component field correlation}
\smallskip

\noindent
The computation is similar in the case of magnetic field with covariance
as in (3.21). One finds:
$$
\eqalignno{
{\bf G}(t)&=\Bigl(0,\ 0,\ {\rm e}^{-2\gamma t}\Bigr)\ , 
&(5.12{\rm a})\cr
{\bf F}(t)&={\rm e}^{-\gamma t}\,\Bigl(\,
(\cosh\delta t+i{\omega\over\delta}\sinh\delta t)\ (1,\ -i,\ 0)\,+\,
{\gamma+2ib\over\delta}\sinh\delta t\ (1,\ i,\ 0)\Bigr)\ ,
&(5.12{\rm b})}
$$
from which one esaily obtains, assuming $\delta>0$:
$$
\eqalign{
&C_t(\vartheta,\varphi;{\bf n})=
{\rm e}^{-2\gamma t}\cos2\vartheta\, n_3\cr
&
-{\rm e}^{-\gamma t}\sin2\vartheta\ 
\Bigl[n_1\Bigl((\cosh\delta t+{\gamma\over\delta}\sinh\delta 
t)\cos\varphi+{\omega+2b\over\delta}\sinh\delta 
t\sin\varphi\Bigr)\cr
&
+n_{2}\Bigl((-\cosh\delta t+{\gamma\over\delta}\sinh\delta 
t)\sin\varphi+{\omega-2b\over\delta}\sinh\delta 
t\cos\varphi\Bigr)\Bigr]\ .
}\eqno(5.13)
$$
The lack of positivity preservation which characterizes 
the time-evolution leading to (5.13) manifests itself in that the 
quantities $C_t(\vartheta,\varphi;{\bf n})$ diverge with large $t$ when 
$\delta>\gamma$.

\bigskip

Expressions (5.9), (5.11) and (5.13) 
agree with those used in [1,2] when there is no 
dissipation, namely putting $a=b=c=\alpha=\beta=\gamma=0$,
$$
C_t(\vartheta,\varphi;{\bf n})=
-n_1\sin2\vartheta\cos(\omega_0t-\varphi)\,-\,
n_2\sin2\vartheta\sin(\omega_0t-\varphi)\,+\, 
n_3\cos2\vartheta\ .
\eqno(5.14)
$$
Notice that the unitary time-evolution generated by
the Hamiltonian $H_0$ contribute to a time varying
redefinition of the angle $\varphi$.

Concerning the issue of
complete positivity vs simple positivity, in  
expressions (5.9) and (5.11) the two possibilities manifest themselves
in different relaxation properties due to whether inequalities
(4.14) or (4.8) are fulfilled.
No physical inconsistencies may affect the
mean values $C_t(\vartheta,\varphi;{\bf n})$;
indeed, negative probabilities may result in negative mean values
of positive observables only if the latter are entangled.
In the case of the quantities involved in inequality (2.10), 
the observables are 
factorized, $P_{1,2}(\vartheta,\varphi)\otimes Q_{{\bf n}}$ 
and the positivity
of their mean values is preserved even when $\Gamma_t$ is only 
positive and not completely positive.

This can be seen as follows.
To the Schr\"odinger time-evolution 
$\rho_{(2)}(t)=({\bf I}_2\otimes\Gamma_t)\,[\rho_{(2)}]$, there
corresponds the Heisenberg time-evolution of
observables $X_{(2)}(t)=({\bf I}_2\otimes 
\Omega_t)\,[X_{(2)}]$,  
$$
{\rm Tr}\Bigl(({\bf I}_2\otimes\Gamma_t)\,[\rho_{(2)}]\,X_{(2)}\Bigr)\,=\,
{\rm Tr}\Bigl(\rho_{(2)}({\bf I}_2\otimes 
\Omega_t)\,[X_{(2)}]\Bigr)\ .
\eqno(5.15)
$$
The maps $\Omega_t$, dual to $\Gamma_t$, form a semigroup
of dynamical maps that transform positive observables into positive 
observables, if the $\Gamma_t$'s preserve the positivity of states. 
Consequently,
even when the initial state $\rho_{(2)}$ is entangled and the $\Gamma_t$'s
positivity preserving, but not completely positive, it turns out that 
$$
{\rm Tr}\Bigl(\rho_{(2)}(t)P_{1,2}(\vartheta,\varphi)
\otimes Q_{{\bf n}}\Bigr)=
{\rm Tr}\Bigl(\rho_{(2)}P_j(\vartheta,\varphi)
\otimes\Omega_t[ Q_{{\bf n}}]\Bigr)\, \geq 0 \ .
\eqno(5.16)
$$

\vskip 1cm

\noindent
{\bf 6. CONCLUSIONS}
\medskip

Complete positivity is a property of quantum time-evolutions  
which is enjoyed by the standard dynamics of closed 
quantum systems generated by Hamiltonian operators, but not 
automatically by the more general reduced dynamics describing time-evolution 
of open quantum systems in interaction with suitable environments.
Complete positivity is intimately related to the phenomenon of quantum 
entanglement, between two different systems, but also between two 
different degrees of freedom of a same physical system.

In this paper we have considered the two entangled degrees of freedom, 
translational and rotational, of a beam of neutrons travelling through
an interferometric apparatus devoted to checking the hypothesis of
noncontextuality.
We have studied the consequences of placing the 
interferometer in a stochastic, gaussian magnetic field weakly
coupled to the spin degree of freedom that provides an experimentally
controllable environment.
As explained
in Section 3,
the same Markov approximation naively yields a
semigroup of dynamical maps ${\bf I}_2\otimes\Gamma_t$, where only
the spin degree of 
freedom evolves in time; by varying the decay properties of the 
external field correlations, these maps turn out to be alternatively
completely positive, simply positivity
preserving, not even positivity preserving.

The noncontextuality tests proposed in [1,2] are based on the 
Clauser-Horne-Shimony-Holt inequality (2.10) without time-dependence,
that is with $t=0$.
The presence of a fluctuating magnetic field induce relaxation on the
spin degree of freedom with strength and properties depending on 
those of the field.
Typically, the mean values in the inequality are damped and make
it more difficult to be violated.
However, in presence of stochastic fields yielding reduced 
dynamics that do not preserve positivity, the inequality might be
dramatically violated because of possible mean values
diverging in time.

This latter possibility is a manifestation of the fact
that any physically consistent
time-evolution $\Gamma_t$ must preserve the positivity of spin states
in order that the eigenvalues of the corresponding spin density matrices
might at any time be used as probabilities, in agreement with the 
statistical interpretation of quantum mechanics.
If the $\Gamma_t$'s  preserve the trace of spin density matrices, but not
their positivity, spin states
may evolve in time in such a way that some of their eigenvalues
become negative, while others greater than $1$, without upper bounds.
It is this physically unacceptable phenomenon that leads to diverging 
mean values.

The request of positivity preservation by the maps $\Gamma_t$ with 
respect to spin states
is thus unexcapable, but
it is not enough to avoid physical inconsistencies when the 
time-evolution maps ${\bf I}_2\otimes\Gamma_t$ act on states 
$\rho_{(2)}$ with correlations between spin and translational degrees 
of freedom.

Inequality (2.10) does reveal the difference between completely positive 
and simply positivity preserving $\Gamma_t$, but only as long as the 
relaxation characteristic are concerned, without any further effect
(as the divergence of some contributions to the inequality).
In fact, the positive observables in (2.10)
are factorized, that is they incorporate no entanglement between the 
translational and spin degree of freedom.
Even if the initial state does incorporate entanglement,
it nevertheless follows that the mean values of factorized observables
remain positive and bounded.

However, the interferometric apparatus proposed in [1,2], 
might also be used to measure the entries of the states of the 
neutron beam at the exit of the interferometer.
In this way, one might have access to the spectrum of an initially
entangled state after being subjected to the effects of the stochastic 
magnetic field.

In the case of fluctuating fields yielding  reduced dynamics 
that preserve positivity, but are not completely positive, 
the theoretical predictions indicate the appearance of negative 
eigenvalues, that is of negative probabilities, in the spectrum of 
the entangled exiting state.
The fact that they are, in line of principle, detectable 
experimentally, does not allow to dismiss such an occurrence as 
practically negligible.
Rather, it forces to reconsider the Markov approximation used to derive 
the time-evolution and to select as physically consistent 
only those providing
completely positive reduced dynamics.

\vfill\eject
 
\noindent
{\bf APPENDIX A}
\medskip

$$
\eqalignno{
\rho_{1{3\atop4}}(t)&={{\cal O}^{1,x}_t(0,0)-{\cal O}_t^{1,-x}(0,0)\over2}
\pm i\,{{\cal O}^{1,-y}_t(0,0)-{\cal O}^{1,y}_t(0,0)\over2}
&(A1)
\cr
\cr
\rho_{2{3\atop4}}(t)&={{\cal O}^{2,x}_t(0,0)-{\cal O}^{2,-x}(0,0)\over2}
\pm i{{\cal O}^{2,-y}_t(0,0)-{\cal O}_t^{2,y}(0,0)\over2}
&(A2)
\cr
\cr
\rho_{{3\atop4}1}(t)&={{\cal O}^{1,z}_t({\pi\over 4},0)-
{\cal O}^{2,z}_t({\pi\over 4},0)\over2}
\mp i{{\cal O}^{1,z}_t({\pi\over 4},-{\pi\over 2})-
{\cal O}^{2,z}_t({\pi\over 4},-{\pi\over 2})\over2}
&(A3)
\cr
\cr
\rho_{{3\atop4}2}(t)&={{\cal O}^{1,-z}_t({\pi\over 4},0)-
{\cal O}^{2,-z}_t({\pi\over 4},0)\over2}
\mp i{{\cal O}^{1,-z}_t({\pi\over 4},-{\pi\over 2})-
{\cal O}^{2,-z}_t({\pi\over 4},-{\pi\over 2})\over2}
&(A4)
\cr
\cr
\rho_{3{3\atop4}}(t)&={{\cal O}^{1,x}_t({\pi\over 4},0)-
{\cal O}^{1,-x}_t({\pi\over 4},0)\over4}
\mp i\,{{\cal O}^{1,y}_t({\pi\over 4},0)-
{\cal O}^{1,-y}_t({\pi\over 4},0)\over4}
\cr
&-{{\cal O}^{2,x}_t({\pi\over 4},0)-
{\cal O}^{2,-x}_t({\pi\over 4},0)\over4}
\pm i\,{{\cal O}^{2,y}_t({\pi\over 4},0)-
{\cal O}^{2,-y}_t({\pi\over 4},0)\over4}
\cr
&-i\,{{\cal O}^{1,x}_t({\pi\over 4},-{\pi\over2})-
{\cal O}^{1,-x}_t({\pi\over 4},-{\pi\over2})\over4}
\mp \,{{\cal O}^{1,y}_t({\pi\over 4},-{\pi\over2})-
{\cal O}^{1,-y}_t({\pi\over 4},-{\pi\over2})\over4}
\cr
&+i\,{{\cal O}^{2,x}_t({\pi\over 4},-{\pi\over2})-
{\cal O}^{2,-x}_t({\pi\over 4},-{\pi\over2})\over4}
\pm \,{{\cal O}^{2,y}_t({\pi\over 4},-{\pi\over2})-
{\cal O}^{2,-y}_t({\pi\over 4},-{\pi\over2})\over4}
&(A5)
\cr
\cr
\rho_{4{3\atop4}}(t)&={{\cal O}^{1,x}_t({\pi\over 4},0)-
{\cal O}^{1,-x}_t({\pi\over 4},0)\over4}
\pm i\,{{\cal O}^{1,y}_t({\pi\over 4},0)-
{\cal O}^{1,-y}_t({\pi\over 4},0)\over4}
\cr
&-{{\cal O}^{2,x}_t({\pi\over 4},0)-
{\cal O}^{2,-x}_t({\pi\over 4},0)\over4}
\mp i\,{{\cal O}^{2,y}_t({\pi\over 4},0)-
{\cal O}^{2,-y}_t({\pi\over 4},0)\over4}
\cr
&+i\,{{\cal O}^{1,x}_t({\pi\over 4},-{\pi\over2})-
{\cal O}^{1,-x}_t({\pi\over 4},-{\pi\over2})\over4}
\mp \,{{\cal O}^{1,y}_t({\pi\over 4},-{\pi\over2})-
{\cal O}^{1,-y}_t({\pi\over 4},-{\pi\over2})\over4}
\cr
&-i\,{{\cal O}^{2,x}_t({\pi\over 4},-{\pi\over2})-
{\cal O}^{2,-x}_t({\pi\over 4},-{\pi\over2})\over4}
\pm \,{{\cal O}^{2,y}_t({\pi\over 4},-{\pi\over2})-
{\cal O}^{2,-y}_t({\pi\over 4},-{\pi\over2})\over4}
&(A6)
}
$$

\vfill\eject

\noindent
{\bf APPENDIX B}
\medskip

After consistent
absorption in the exponentials of terms linear in $t$,
the entries ${\cal G}_{ij}(t)$, $i,j=1,2,3$,
of the matrix ${\cal G}_t$ solution of equation (4.5),
calculated up to first order in the dissipative term $\cal D$, 
can be expressed as
$$
\eqalignno{
{\cal G}_{11}(t)&={\rm e}^{-(a+\alpha)t}\cos\omega_0t\,+\,
{\alpha-a\over\omega_0}\sin\omega_0t\ ,
&(B1)\cr
{\cal G}_{12}(t)&=-\Bigl({\rm e}^{-(a+\alpha)t}\,+\,
{2b\over\omega_0}\Bigl)\sin\omega_0t\ ,
&(B2)\cr
{\cal G}_{13}(t)&=-{4\over\omega_0}\sin{\omega_0\over2}t\,
\Bigl(c\,\cos{\omega_0\over2}t\,-\,\beta\sin{\omega_0\over 2}t\Bigr)\ ;
&(B3)}
$$
$$
\eqalignno{
{\cal G}_{21}(t)&=\Bigl({\rm e}^{-(a+\alpha)t}\,-\,{2b\over\omega_0}\Bigr)\,
\sin\omega_0t\ ,
&(B4)\cr
{\cal G}_{22}(t)&={\rm e}^{-(a+\alpha)t}\cos\omega_0t\,+\,
{a-\alpha\over\omega_0}\sin\omega_0t\ ,
&(B5)\cr
{\cal G}_{23}(t)&=-{4\over\omega_0}\sin{\omega_0\over2}t\,
\Bigl(\beta\,\cos{\omega_0\over2}t\,+\,c\sin{\omega_0\over 2}t\Bigr)\ ;
&(B6)}
$$
$$
\eqalignno{
{\cal G}_{31}(t)&=-{4\over\omega_0}\sin{\omega_0\over2}t\,
\Bigl(c\,\cos{\omega_0\over2}t\,+\,\beta\sin{\omega_0\over 2}t\Bigr)
&(B7)\cr
{\cal G}_{32}(t)&=
{4\over\omega_0}\sin{\omega_0\over2}t\,
\Bigl(c\,\sin{\omega_0\over2}t\,-\,\beta\cos{\omega_0\over 2}t\Bigr)
&(B8)\cr
{\cal G}_{33}(t)&={\rm e}^{-2\gamma t}\ .
&(B9)}
$$

\vfill\eject

\centerline{\bf REFERENCES}
\vskip 1cm
\item{1.} S. Basu, S. Bandyopadhyay, G. Kar and D. Home, Phys. Lett. {\bf A 279}
(2001) 281
\smallskip
\item{2.} M. Michler, H. Weinfurter and M. $\dot{\hbox{Z}}$ukowski, 
Phys. Rev. Lett. {\bf 84} (2000) 5457
\smallskip
\item{3.} J.F. Clauser, M.A. Horne, A. Shimony and R.A. Holt, Phys. Rev. Lett. 
{\bf 23} (1969) 880
\smallskip
\item{4.} A. Datta and D. Home, Found. Phys. Lett. {\bf 4} (1991) 165
\smallskip
\item{5.} G. Ghirardi, R. Grassi and R. Ragazzon, in The Da$\Phi$ne Physics
Handbook, Vol. I, L. Maiani, G. Pancheri and N. Paver eds., (INFN, Frascati, 
1992)
\smallskip
\item{6.} A. Di Domenico, Nucl. Phys. {\bf B450} (1995) 293
\smallskip
\item{7.} E.B. Davies, {\it Quantum theory of Open systems} 
(Academic Press, London, 1976)
\smallskip
\item{8.}
V. Gorini, A. Frigerio, M. Verri, A. Kossakowski and E.G.C. Sudarshan,
Rep. Math. Phys. {\bf 13} (1978) 149
\smallskip
\item{9.} H. Spohn, Rev. Mod. Phys. {\bf 52} (1980) 569
\smallskip
\item{10.} R. Alicki and K. Lendi, {\it Quantum Dynamical Semigroups and
Applications}, Lect. Notes Phys. {\bf 286}, (Springer-Verlag, Berlin, 1987)
\smallskip
\item{11.} K. Kraus, Ann. Phys. {\bf 64} (1971) 311 
\smallskip
\item{12.}
M. Choi, Linear Alg. Appl. {\bf 10} (1975) 285 
\smallskip
\item{13.}
J. Budimir and J.L. Skinner, J. Stat. Phys. {\bf 49} (1987) 1029
\smallskip
\item{14.}
B.B. Laird and J.L. Skinner, J. Chem. phys. {\bf 94} (1991) 4405
\smallskip
\item{15.}
A. Suarez, R. Silbey and I. Oppenheim, J. Chem. Phys. {\bf 97} (1992) 5101
\smallskip
\item{16.}
P. Pechukas, Phys. Rev. Lett. {\bf 73} (1994) 1060
\smallskip
\item{17.}
R. D\"umcke and H. Spohn, Z. Physik {\bf B34} (1979) 419
\smallskip
\item{18.}
F. Benatti and R. Floreanini, Nucl. Phys. {\bf B488} (1997) 335
\smallskip
\item{19.}
F. Benatti and R. Floreanini, Nucl. Phys. {\bf B401} (1998) 550
\smallskip
\item{20.}
F. Benatti and R. Floreanini, Phys. Lett. {\bf B468} (1999) 287
\smallskip
\item{21.}
F. Benatti, R. Floreanini and R. Romano, Nucl. Phys. {\bf B602} (2001) 541
\smallskip
\item{22.}
V. Gorini and A. Kossakowski, J. Math. Phys. {\bf 17} (1976) 1298
\smallskip
\item{23.} J.L. Staudenmann, S.A. Werner, R. Colella and A.W. Overhauser,
Phys. Rev. A {\bf 21} (1980) 1419
\smallskip
\item{24.} S.A. Werner and A.G. Klein, Meth. Exp. Phys. {\bf A23} (1986) 259
\smallskip
\item{25.} V.F. Sears, {\it Neutron Optics}, (Oxford University Press, Oxford,
1989)
\smallskip
\item{26.} {\it Advance in Neutron Optics and Related Research Facilities},
M. Utsuro, S. Kawano, T. Kawai and A. Kawaguchi, eds.,
J. Phys. Soc. Jap. {\bf 65}, Suppl. A, 1996
\smallskip
\item{27.} K.C. Littrell, B.E. Allman and S.A. Werner,
Phys. Rev. A {\bf 56} (1997) 1767
\smallskip
\item{28.} B.E. Allman, H. Kaiser, S.A. Werner, A.G. Wagh, V.C. Rakhecha
and J. Summhammer, Phys. Rev A {\bf 56} (1997) 4420
\smallskip
\item{29.} H. Rauch, S. A. Werner, {\it Neutron Interferometry} (Oxford
University Press, Oxford 2000)
\smallskip
\item{30.}
G. Lindblad, Comm. Math. Phys. {\bf 48} (1976) 119
\smallskip
\item{31.}
A. Gorini, A, Kossakowski and E.C.G. Sudarshan, J. Math. Phys. {\bf 17} (1976) 
821
\smallskip
\item{32.}
C.P. Slichter, {\it Principles of Magnetic Resonance}
 (Springer Verlag, Berlin, 1990)
\bye